\documentclass[aps,prl,twocolumn,superscriptaddress,groupedaddress]{revtex4}  

\usepackage{graphicx}  
\usepackage{dcolumn}   
\usepackage{pbox}
\usepackage{bm}        
\usepackage{amssymb}   
\usepackage{amsmath}
\usepackage{siunitx}
\usepackage{xfrac}
\usepackage{verbatim}
\usepackage{doi}
\usepackage{filecontents}

\newcommand{\eref}[1]{Eq.~(\ref{#1})}
\newcommand{\tref}[1]{Table~\ref{#1}}

\newcommand{\ket}[1]{\left|#1\right\rangle}

\begin{document}

\title{Enhancing Divalent Optical Atomic Clocks with the $^{1}\mathrm{S}_0$$\leftrightarrow$$^{3}\mathrm{P}_{2}$ Transition}


\author{Matthew A. Bohman}
\affiliation{HEP Division, Argonne National Laboratory, Lemont Illinois, 60439, USA}
\affiliation{Time and Frequency Division, National Institute of Standards and Technology, Boulder, Colorado, 80305, USA}

\author{Sergey G. Porsev}
\affiliation{Department of Physics and Astronomy, University of Delaware, Newark, Delaware 19716, USA}

\author{David B. Hume}
\affiliation{Time and Frequency Division, National Institute of Standards and Technology, Boulder, Colorado, 80305, USA}

\author{David R. Leibrandt}
\affiliation{Time and Frequency Division, National Institute of Standards and Technology, Boulder, Colorado, 80305, USA}
\affiliation{Department of Physics and Astronomy, University of California, Los Angeles, CA, USA}
\affiliation{Department of Physics, University of Colorado, Boulder, Colorado, 80309, USA}

\author{Marianna S. Safronova}
\affiliation{Department of Physics and Astronomy, University of Delaware, Newark, Delaware 19716, USA}

\date{\today}

\begin{abstract} 
Divalent atoms and ions with a singlet $S$ ground state and triplet $P$ excited state form the basis of many high-precision optical atomic clocks. Along with the metastable $^{3}\mathrm{P}_{0}$ clock state, these atomic systems also have a nearby metastable $^{3}\mathrm{P}_{2}$ state. We investigate the properties of the electric quadrupole $^{1}\mathrm{S}_0$$\leftrightarrow$$^{3}\mathrm{P}_{2}$ transition with a focus on enhancing already existing optical atomic clocks. In particular, we investigate the $^{1}\mathrm{S}_0$$\leftrightarrow$$^{3}\mathrm{P}_{2}$ transition in $^{27}\mathrm{Al}^{+}$ and calculate the differential polarizability, hyperfine effects, and other relevant atomic properties. We also discuss potential applications of this transition, notably that it provides two transitions with different sensitivities to systematic effects in the same species. In addition, we describe how the $^{1}\mathrm{S}_0$$\leftrightarrow$$^{3}\mathrm{P}_{2}$ transition can be used to search for physics beyond the Standard Model and motivate investigation of this transition in other existing optical atomic clocks.  
\end{abstract}

\maketitle

\section{Introduction}
Optical atomic clock species with two long-lived excited states have proven to be very useful in constraining systematic effects and searching for new physics beyond the Standard Model. Prominently, the most stringent constraints on the time variation of the fine structure constant come from a comparison of two clock transitions in Yb$^{+}$ \cite{Lange_alphadot, Yb+_new} - an electric quadrupole transition (E2), and electric octupole transition (E3). In addition, while the E3 transition is much narrower and is more suited to low uncertainty, high stability clock operation, the E2 transition provides important information on systematic effects, for example, allowing in situ measurement of the electric quadrupole field \cite{Huntemann10-18}. Here, we consider similar applications with the previously unused $^{1}\mathrm{S}_0$$\leftrightarrow$$^{3}\mathrm{P}_{2}$ transition in divalent systems using $^{27}$Al$^+$ as a specific example. We find that the $^{1}\mathrm{S}_0$$\leftrightarrow$$^{3}\mathrm{P}_{2}$ has uses as an auxiliary transition which, when used along with the well established $^{1}\mathrm{S}_0$$\leftrightarrow$$^{3}\mathrm{P}_{0}$ transition, can provide insights into atomic structure calculations, inform clock-related systematic shifts on both transitions, and provide higher measurement stability due to the longer excited state lifetime. Although the $^{1}\mathrm{S}_0$$\leftrightarrow$$^{3}\mathrm{P}_{2}$ transition involves a $J \neq 0$ state and requires a slightly different spectroscopy sequence, the transition retains the insensitivity to blackbody radiation of the $^{1}\mathrm{S}_0$$\leftrightarrow$$^{3}\mathrm{P}_{0}$ transitions and is amenable to high-accuracy clock operation.

\section{Divalent Atomic Clock Species}
Common species of optical atomic clocks are divalent atoms or ions, with a $^{1}\mathrm{S}_0$ ground state and a metastable $^{3}\mathrm{P}_{0}$ excited state. Among the most prominent examples are neutral Sr and Yb and singly-ionized Al$^+$ and In$^+$ \cite{Optical_atomic_clocks_review}. Illustrated in Fig$.$\,1\,a$)$, the ions are confined in a linear rf trap and the neutral atoms in an optical lattice. For each of these species, a nearby $^{1}\mathrm{S}_0 \leftrightarrow \, ^{3}\mathrm{P}_{1}$ transition, which has a lifetime many orders of magnitude shorter than the strictly forbidden clock transition, provides narrow-line cooling, efficient state preparation, and readout. A dipole-allowed $^{1}\mathrm{S}_0 \leftrightarrow \, ^{1}\mathrm{P}_{1}$ transition also allows for state detection and broader Doppler cooling, although in the case of ions this transition lies in the vacuum ultra-violet (VUV) and is unused. As a result, these clocks use co-trapped ions with more easily accessible cooling transitions for sympathetic cooling and state readout via quantum logic spectroscopy \cite{Schmidt_QLS}. However, in all these systems, there also exists an additional metastable $^{3}\mathrm{P}_{2}$ state that has typically not been utilized, although in some neutral atoms this transitions has been considered for applications with many-body physics and quantum simulation \cite{Sr_3P2, Yb_3P2}. Here, we investigate the clock-related properties of the $^{1}\mathrm{S}_0 \leftrightarrow \, ^{3}\mathrm{P}_{2}$ transition, and, for concreteness, focus on $^{27}$Al$^+$. While many of our results inspire further investigation with other systems, the case of $^{27}$Al$^+$ was particularly interesting, as clock comparison measurements with existing $^{27}$Al$^+$ ion clocks are close to being limited by the \SI{20.7}{\second} lifetime of the $^{3}\mathrm{P}_{0}$ state \cite{Cle20}. The longer-lived $^{3}\mathrm{P}_{2}$ state with a lifetime of around \SI{300}{\second} is occasionally populated by background gas collisions during typical clock operation and has long been considered as a possible means of obtaining higher stability in the same single ion system. In addition, the $^{3}\mathrm{P}_{2}$ state lies only around \SI{1}{\tera\hertz} higher than the $^{1}\mathrm{S}_0$$\leftrightarrow$$^{3}\mathrm{P}_{0}$ transition and could be driven with minimal changes to the clock laser.\\

\begin{figure*}
    \centering
    \includegraphics[width=\linewidth]{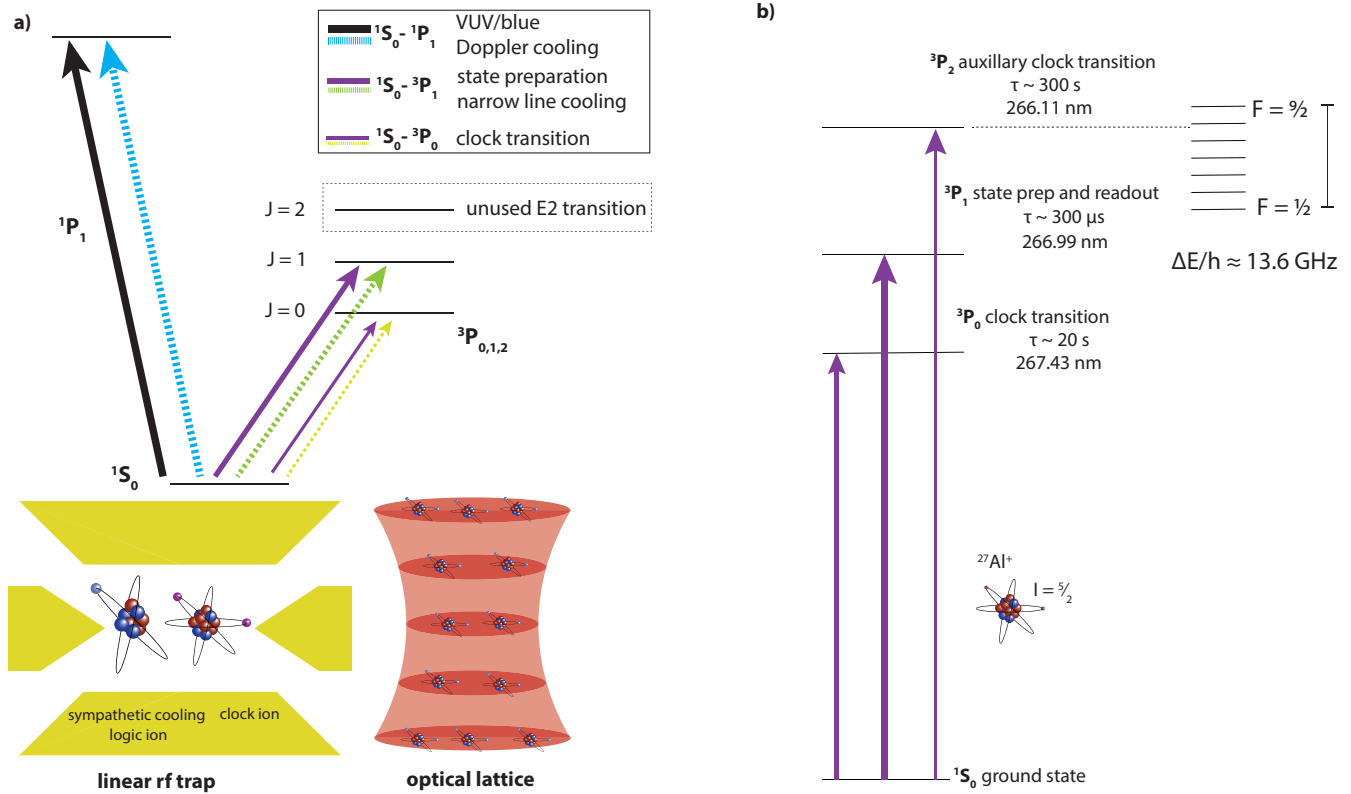}
    \caption{\textbf{a)} The low-lying energy levels of divalent ions (left) and neutral atoms (right) are shown, not to scale. Neutral atoms can be confined in an optical lattice while ions can be stored in a linear rf trap, enabling the use of an additional ion for sympathetic cooling or quantum logic. Additional low-lying D-levels, not shown, exist in divalent neutral atoms but are at much higher energy in ions. \textbf{b)} The level structure of $^{27}$Al$^+$ is shown, including the hyperfine splitting of the $^{3}\mathrm{P}_{2}$ state. The line thickness illustrates the relative strength, not to scale, of the transitions.}
    \label{fig:Al+ levels}
\end{figure*}

More generally, we emphasize the utility of an additional narrow linewidth transition in a well-established atomic species. In the case of trapped ions, motional shifts, which can constitute a large fraction of the uncertainty budget, are exactly the same for both transitions. As a result, differential frequency measurements between the two transitions, either with co-trapped ions \cite{Chou_3ion} or a single ion, can reach a higher precision than absolute frequency measurements or frequency ratio measurements between two completely different systems. In addition, some optical atomic clock species, notably Yb$^{+}$ and Lu$^{+}$, routinely rely on auxiliary transitions to constrain shifts of the primary clock transition using only one ion \cite{Huntemann10-18, Lu+2022comparison}. Furthermore, in the case of the Yb$^{+}$ ion clock, high-precision frequency ratio measurements between the E2 and E3 transitions provide extremely tight constraints on certain types of proposed physics beyond the Standard Model \cite{Lange_alphadot}. Despite the sparse energy level spacing and lack of a level crossing in divalent ions, the existence of long lived $^{3}\mathrm{P}_{0}$ and $^{3}\mathrm{P}_{2}$ states, in principle, allows for similar measurements.\\

In the following sections, we first detail calculations of the clock-related atomic properties of the $^{1}\mathrm{S}_0$$\leftrightarrow$$^{3}\mathrm{P}_{2}$ transition in  $^{27}$Al$^+$. We then describe averaging schemes that have been developed for other species with $J\neq0$ and could be applied to the $^{1}\mathrm{S}_0$$\leftrightarrow$$^{3}\mathrm{P}_{2}$ transition in $^{27}$Al$^+$ to compensate leading systematic shifts. We also discuss possible applications of the $^{1}\mathrm{S}_0$$\leftrightarrow$$^{3}\mathrm{P}_{2}$ transition in  $^{27}$Al$^+$. Of particular note, atomic structure calculations are available for both clock transitions in $^{27}$Al$^+$ and an accurate frequency ratio measurement of the $^{1}\mathrm{S}_0 \leftrightarrow \, ^{3}\mathrm{P}_{0}$ and $^{1}\mathrm{S}_0 \leftrightarrow \, ^{3}\mathrm{P}_{2}$ transitions can provide much needed theoretical input. Finally, we identify potential measurements of physics beyond the Standard Model and a measurement of the blackbody radiation environment that, while challenging for the specific case of $^{27}$Al$^+$, are likely to be of interest for similar species.\\


\section{Atomic Properties}
\subsection{Method of calculation}
To calculate systematic effects of this new clock transition, we first start with a general description of atomic properties and perform theoretical calculations of $^{27}$Al$^+$. As discussed above, $^{27}$Al$^+$ is a divalent ion and contains $[1s^2, 2s^2, 2p^6]$ core electrons and two valence electrons. We use a relativistic high-precision approach that combines configuration interaction (CI) and coupled-cluster (CC) methods \cite{DzuFlaKoz96,SafKozJoh09} that allows us to efficiently include correlation corrections.\\

We start with a solution of the Dirac-Hartree-Fock (DHF) equations to construct the finite basis set used in the calculations. The $3s,4s$, $3p,4p$, $3d$, and $4d$ orbitals were constructed in the DHF Al$^{3+}$ frozen core potential. The remaining virtual orbitals were formed using a recurrent procedure described in Refs$.$\,\cite{KozPorFla96,KozPorSaf15} and the newly constructed functions were then orthonormalized with respect to the functions of the same symmetry. The basis sets included a total of six partial waves ($l_{\rm max} = 5$) and orbitals with a principal quantum number $n$ up to 25. We included the Breit interaction on the same basis as the Coulomb interaction in the stage of constructing the basis set.\\

In a CI+CC approach  that allows us to include core-valence correlations~\cite{DzuFlaKoz96,SafKozJoh09},
the wave functions and energy levels of the valence electrons were found by solving the multiparticle relativistic equation~\cite{DzuFlaKoz96},
\begin{equation}
H_{\rm eff}(E_n) \Phi_n = E_n \Phi_n,
\label{Heff}
\end{equation}
where the effective Hamiltonian is defined as
\begin{equation}
H_{\rm eff}(E) = H_{\rm FC} + \Sigma(E),
\label{Heff1}
\end{equation}
where $H_{\rm FC}$ is the Hamiltonian in the frozen-core approximation.
The energy-dependent operator $\Sigma(E)$ accounts for the virtual excitations of the core electrons and is constructed in two ways: using (i) second-order many-body perturbation theory (MBPT) over the residual Coulomb interaction~\cite{DzuFlaKoz96}
and (ii) the linearized coupled-cluster single-double (LCCSD) method~\cite{SafKozJoh09}. We refer to these approaches as the CI + MBPT and CI + all-order methods and treat the difference between the two results as
the uncertainty of our calculation.

\subsection{Energy levels and Polarizabilities}
To calculate the differential polarizabilities of the two clock states, we begin by calculating the low-lying energy levels in the pure CI, CI+MBPT, and CI+all-order approximations. To verify the convergence of the CI approach, we performed several calculations and sequentially increased the size of the configuration space. Sets of configurations were constructed by including single and double excitations from the main even ($3s^2$) and odd ($3s3p$) configurations in the upper shells. The set of configurations in which the convergence of CI was achieved included excitations to the orbitals up to $20s,20p,20d,20f,20g$, designated $[20spdfg]$. The results of these calculations are presented in \tref{Tab:E}. For the $3s^2\, ^1\!S_0$ state we present the valence energy (in a.u.) which can be compared to the sum of two ionization potentials IP(Al$^+$) and IP(Al$^{2+}$)~\cite{RalKraRea11}.\\

\begin{table*}[t]
\caption{The low-lying energies (in cm$^{-1}$) of the even- and odd-parity levels
calculated in the pure CI (labeled as ``CI''), CI+MBPT (labeled as ``CI+MBPT'') and CI+all-order (labeled as ``CI+All'') approximations.
For the $3s^2\, ^1\!S_0$ state, we present its valence energy (in a.u.) which can be compared to the sum of two ionization potentials
IP(Al$^+$) and IP(Al$^{2+}$)~\cite{RalKraRea11}. The energies of the excited states (given in cm$^{-1}$) are counted from the ground state. The experimentally determined values (labeled as ``NIST Exp.'') are taken from the NIST Atomic Spectra Database \cite{RalKraRea11} and the differences between theoretical and experimental values are given in the following columns, labeled ``NIST-CI'', ``NIST-MBPT'', and ``NIST-All'', respectively.}
\label{Tab:E}%
\begin{ruledtabular}
\begin{tabular}{lccccccc}
                 &   CI      &  CI+MBPT  &  CI+All  & NIST Exp.~\cite{RalKraRea11} & NIST-CI & NIST-MBPT  & NIST-All\\
\hline \\[-0.6pc]
$3s^2\, ^1\!S_0$ & 1.71583   & 1.73640   & 1.73712  &  1.73737                & 1.2\%   & 0.06\%     &  0.01\% \\[0.2pc]
$3p^2\, ^1\!D_2$ &  83508    &  85423    &  85509   &  85481                  & 2.3\%   & 0.07\%     & -0.03\% \\[0.2pc]
$3s4s\, ^3\!S_1$ &  90106    &  91316    &  91370   &  91275                  & 1.3\%   &-0.05\%     & -0.10\% \\[0.2pc]
$3p^2\, ^3\!P_0$ &  92598    &  94062    &  94095   &  94085                  & 1.6\%   & 0.02\%     & -0.01\% \\[0.2pc]
$3p^2\, ^3\!P_1$ &  92657    &  94127    &  94158   &  94147                  & 1.6\%   & 0.02\%     & -0.01\% \\[0.2pc]
$3p^2\, ^3\!P_2$ &  92772    &  94247    &  94279   &  94269                  & 1.6\%   & 0.02\%     & -0.01\% \\[0.2pc]
\hline  \\[-0.6pc]
$3s3p\, ^3\!P_0$ &  55236    &  54083    &  53548   &  53548                  & 3.0\%   &  0.09\%    & -0.06\%  \\[0.2pc]
$3s3p\, ^3\!P_1$ &  56858    &  55707    &  55196   &  55196                  & 3.0\%   &  0.09\%    & -0.06\%  \\[0.2pc]
$3s3p\, ^3\!P_2$ &  60979    &  59757    &  59228   &  59228                  & 3.0\%   &  0.09\%    & -0.06\%  \\[0.2pc]
$3s3p\, ^1\!P_1$ &  80162    &  80166    &  79912   &  79912                  & 0.5\%   & -0.06\%    & -0.03\%  
\end{tabular}
\end{ruledtabular}
\end{table*}
We then found the static polarizabilities of states $3s^2\,^1\!S_0$, $3s3p\,^3\!P_0$, and $3s3p\,^3\!P_2$
and the differential polarizabilities $\alpha(^3\!P_0) -\alpha(^1\!S_0)$ and $\alpha(^3\!P_2) -\alpha(^1\!S_0)$ with the scalar and tensor components of the total $^3\!P_2$ polarizability defined as
\begin{eqnarray}
\alpha (^3\!P_2,m) = \alpha_0 (^3\!P_2) + \alpha_2 (^3\!P_2) \left(\frac{m^2}{2} - 1 \right).
\label{a1}
\end{eqnarray}
where $m$ is the projection of the total angular momentum $J=2$. The uncertainties of $\Delta \alpha$ are given by the difference between the CI+all-order and CI+MBPT values, both in \tref{Tab:Polar}.\\

As seen in \tref{Tab:Polar}, the polarizabilities of the $^{1}\mathrm{S}_0 \leftrightarrow$$^{3}\mathrm{P}_{0}$ and $^{1}\mathrm{S}_0\leftrightarrow$$^{3}\mathrm{P}_{2}$ states are quite similar, with the differential polarizability of the $^{1}\mathrm{S}_0 \leftrightarrow \, ^{3}\mathrm{P}_{2}$ transition being slightly smaller. As a result, the $^{1}\mathrm{S}_0 \leftrightarrow \, ^{3}\mathrm{P}_{2}$ transition will have a sensitivity to blackbody radiation (BBR) similar to the $^{1}\mathrm{S}_0 \leftrightarrow \, ^{3}\mathrm{P}_{0}$ transition, which is among the lowest of all currently existing clock species. In a later section, we describe how the existence of two narrow clock transitions in the same species can provide an in situ measurement of the BBR environment, although for the case of $^{27}$Al$^+$ this measurement is challenging, as the fractional BBR shift at room temperature of both transitions is on the order of \SI{1e-18}{}. We note, however, that the calculations of the differential polarizabilities of both transitions are very similar, and, as a result, measurement of both differential polarizabilities would provide valuable insight into the accuracy of these calculations. Measurement schemes have been demonstrated that enable measurement of the differential polarizability of a clock transition to high accuracy \cite{Barrett_polarizability, Lu+2022comparison}, potentially benchmarking atomic theory in mid-$Z$ atoms.

\begin{table}[t]
\caption{The static polarizabilities of the $^1\!S_0$, $^3\!P_0$, and $^3\!P_2$ states (including the core and cv contributions)
and the differential polarizabilities $\Delta \alpha$ (in $a_0^3$, where $a_0$ is the Bohr radius), calculated in the pure CI, CI+MBPT and CI+all-order approximations,
are presented. The total polarizability $\alpha (^3\!P_2)$ was calculated at $m = 1$. The uncertainties are given in parentheses.}
\label{Tab:Polar}%
\begin{ruledtabular}
\begin{tabular}{lccccc}
         &                               &     CI     &  CI+MBPT   &   CI+All     & Ref.~\cite{SafKozCla11} \\
\hline \\ [-0.6pc]
 Static  & $\alpha(^1\!S_0)$             &  24.449    &   24.091   &  24.096      &  24.048      \\[0.2pc]
         & $\alpha(^3\!P_0)$             &  24.907    &   24.567   &  24.582      &  24.543     \\[0.2pc]
         & $\Delta \alpha$               &   0.458    &    0.476   &  $0.486(10)$ &   0.495     \\[0.5pc]
         & $\alpha_0(^3\!P_2)$           &  25.016    &   24.679   &  24.695      &             \\[0.2pc]
         & $\alpha_2(^3\!P_2)$           &   0.610    &    0.562   &   0.565      &             \\[0.2pc]
         & $\alpha(^3\!P_2)$             &  24.711    &   24.398   &  24.413      &             \\[0.2pc]
         & $\Delta \alpha$               &   0.262    &    0.307   &  $0.317(10)$ &
\end{tabular}
\end{ruledtabular}
\end{table}

\subsection{Zeeman Shift}
The linear Zeeman shift of the $^{1}\mathrm{S}_0 \leftrightarrow \, ^{3}\mathrm{P}_{2}$ transition is larger than that of the $^{1}\mathrm{S}_0 \leftrightarrow \, ^{3}\mathrm{P}_{0}$ transition by roughly $\mu_B/\mu_N$, the ratio of the Bohr magneton to the nuclear magneton. In addition, there exists a quadratic Zeeman shift, including a small component that is proportional to the magnetic dipole polarizability of the $^{3}\mathrm{P}_{2}$ state that arises from mixing with the nearby $^{3}\mathrm{P}_{1}$ state. Although, in general, averaging over Zeeman sublevels is necessary for high accuracy clock operation, as with the $^{1}\mathrm{S}_0 \leftrightarrow \, ^{3}\mathrm{P}_{0}$ \cite{Bre18} transition, we first calculate the hyperfine constants and the Zeeman shifts for a fixed sublevel.
\subsubsection{Hyperfine Constants and Zeeman Shift}
We first calculated the magnetic dipole and electric quadrupole hyperfine constants $A_{\mathrm{hfs}}$ and $B_{\mathrm{hfs}}$ for the $^3\!P_2$ state. For $^{27}$Al$^+$ with nuclear spin $I=5/2$, the nuclear magnetic moment $\mu_I = 3.6415069(7)\,\mu_N$~\cite{Sto05} and the nuclear quadrupole moment, $Q_{\rm nuc} = 0.1466(6)\, {\rm b}$~\cite{Pyy18}.\\

In the presence of a weak external magnetic field $\bf B$, we need to consider both the hyperfine and Zeeman interactions:
\begin{equation}
H = H_{\rm hfs} - {\boldsymbol \mu}_{\rm at} {\bf B}
\end{equation}
where ${\boldsymbol \mu}_{\rm at} = -\mu_0 g_J {\bf J} - \mu_N g_I {\bf I}$.
Here, $g_J$ is the electron $g$-factor, given in the non-relativistic approximation by the formula
\begin{equation}
g_J = \frac{3}{2} + \frac{S(S+1) - L(L+1)}{2J(J+1)}
\label{g_J}
\end{equation}
and $g_I = \mu_I/(I\,\mu_N) \approx 1.4566$. In the absence of an external magnetic field, the hyperfine splitting is given by
\begin{eqnarray}
\Delta E_{\rm hfs}/h &\equiv& \left(\Delta E^{(1)}_{\rm hfs} + \Delta E^{(2)}_{\rm hfs} \right)/h  \\
                     & = & \frac{1}{2} A_{\mathrm{hfs}} K +\\
                     &  & B_{\mathrm{hfs}}\, \frac{3/4K(K+1)-J(J+1)I(I+1)}{2I(2I-1)J(2J-1)}, \nonumber
\end{eqnarray}
where
$K \equiv F(F+1) - J(J+1) - I(I+1)$ and $h$ is the Planck constant.\\

Our calculation within the framework of the CI + all-order approximation (including random phase approximation corrections) gives $A_{\mathrm{hfs}} \approx 1132\,{\rm MHz}$ and $B_{\mathrm{hfs}} \approx 30\,{\rm MHz}$. As an example, for the total momentum $F=1/2$, we find for the $^3\!P_2$ state
\begin{eqnarray}
\Delta E^{(1)}_{\rm hfs}/ h &\approx& -7924\,\, {\rm MHz}, \nonumber \\
\Delta E^{(2)}_{\rm hfs}/h &\approx&    21\,\, {\rm MHz} .
\end{eqnarray}

 We then estimate the quadratic Zeeman shift without any averaging and neglect the contribution of the electric-quadrupole interaction to the hyperfine splitting, as it is 400 times smaller than the contribution of the magnetic-dipole interaction.\\

The operator $H_{\rm hfs}$ is diagonal in both $F$ and $M$ (where $M$ is the projection of ${\bf F}$) while the operator ${\boldsymbol \mu}_{\rm at}$ is diagonal in $M$ but not in $F$. To take into account hyperfine and Zeeman interactions, we use a basis of
$|Jm,I\lambda \rangle$ (where $\lambda$ is the projection of $I$) or just $|m,\lambda\rangle$ since $J$ and $I$ are constants within a given level.
If $M= m + \lambda$ is fixed, we have the basis $|m, M-m \rangle$. In the case of a weak magnetic field, the Zeeman interaction can be treated as a perturbation to the $|F,M \rangle$ basis.\\

Although it is generally difficult to isolate the quadratic component of the Zeeman shift, we consider the extreme state $F=9/2$ and denote the magnitude of the applied magnetic field as $B$. It can be shown (see the Appendix for more details) that for $m=2$, $\lambda=5/2$, and $M=9/2$ the 
matrix element is
\begin{eqnarray}
\langle m&=&2, \lambda=5/2 \,| H |\, m=2, \lambda=5/2 \rangle \nonumber \\
&=& 5hA_{\mathrm{hfs}} + 2\mu_0 g_J B
\end{eqnarray}
 and does not contain the term $\sim B^2$. For $M=7/2$ there are only two possible states $|m=2, \lambda=3/2 \rangle$ and $|m=1, \lambda=5/2 \rangle$
and we can define them as the basis. The quadratic contribution in $B$, designated as $\Delta E^{(2)}$,
is (see Appendix)
\begin{eqnarray}
\left|\Delta E^{(2)}\right| = \frac{1}{18}\,\frac{(\mu_0 g_J)^2}{h A_{\mathrm{hfs}}}\,B^2 .
\label{x2}
\end{eqnarray}

\noindent Using $g_J(^3\!P_2) = 3/2$ and $A_{\mathrm{hfs}}(^3\!P_2) =  1132\,{\rm MHz}$, we find that for $^3\!P_2$,
\begin{eqnarray}
\left|\Delta E^{(2)}\right| &\approx& 1.433 \times 10^{-23}\, \frac{{\rm J}}{{\rm T}^2}\,  B^2 ,  \\
\Delta \nu^{(2)} \equiv \Delta E^{(2)}/h   &\approx& 22\, \frac{{\rm kHz}}{({\rm mT})^2} B^2.
\end{eqnarray}
\noindent We again emphasize that while these shifts are large, they are easily taken into account by proper averaging over the Zeeman sublevels. As these averaging schemes suppress not only Zeeman shifts but additional higher order field shifts such as the electric quadrupole shift, we discuss the details in a later section.

\subsubsection{Contribution proportional to $M1$ polarizability}
Nevertheless, a small residual quadratic Zeeman shift due to the magnetic dipole polarizability of the $^3\!P_2$ state remains, even with proper hyperfine averaging. This shift is roughly inversely proportional to the fine structure splitting of the $^3P$ manifold and gives rise to the energy splitting, $\Delta E$ ~\cite{PorSaf20}
\begin{eqnarray}
 \Delta E = - \frac{1}{2} \alpha^{\rm M1} B^2 .
\label{DelE}
\end{eqnarray}

\noindent The magnetic dipole polarizability $\alpha^{M1}$ is given by
\begin{eqnarray}
\alpha^{\rm M1} =  \alpha_0^{\rm M1} + \alpha_2^{\rm M1}\, \frac{3m^2-J(J+1)}{J(2J-1)} ,
\label{aM1}
\end{eqnarray}
where $\alpha_0^{\rm M1}$ and $\alpha_2^{\rm M1}$ are the scalar and tensor parts of the
magnetic dipole polarizability.\\


For the $^3\!P_2$ state, the scalar static and tensor polarizabilities are given by
\begin{eqnarray}
\alpha_0^{\rm M1} &=& \frac{2}{15}\, \sum_n \frac{|\langle n ||\mu||^3\!P_2 \rangle|^2}{E(n) - E(^3\!P_2)} \nonumber \\
\alpha_2^{\rm M1} &=& \frac{4}{\sqrt{21}} \nonumber \\
&\times& \sum_{n} (-1)^{J_n}
\left\{
\begin{array}{ccc}
2 & 1 & J_n \\
1 & 2 & 2
\end{array}
\right\}
\frac{|\langle n ||\mu|| ^3\!P_2 \rangle|^2} {E(n) - E(^3\!P_2)},
\label{alphaM1}
\end{eqnarray}
where $\bm \mu$ is the magnetic dipole moment operator.\\

To estimate the shift for the clock transition due to this term, we note that the 
$\alpha^{M1}(^1\!S_0)$  polarizability is negligibly small
compared to $\alpha^{M1}(^3\!P_2)$, so we have
\begin{equation*}
\Delta \nu^{(1)} \equiv \frac{\Delta E(^3\!P_2) - \Delta E(^1\!S_0)}{h} \approx \Delta E(^3\!P_2)/h .
\end{equation*}

\noindent For an estimate of $\alpha^{M1}(^3\!P_2)$ we take into account that the main contribution to this polarizability comes from the intermediate
state $3s3p\,\,^3\!P_1$. Then, using \eref{alphaM1}, we obtain
\begin{eqnarray}
\alpha_0^{\rm M1}(^3\!P_2) &\approx&  \frac{2}{15}\, \frac{\langle ^3\!P_1 || \mu || ^3\!P_2 \rangle^2}{E(^3\!P_1) - E(^3\!P_2)} , \nonumber \\
\alpha_2^{\rm M1}(^3\!P_2) &\approx& -\frac{2}{15}\, \frac{\langle ^3\!P_1 || \mu || ^3\!P_2 \rangle^2}{E(^3\!P_1) - E(^3\!P_2)} .
\end{eqnarray}

\noindent In this approximation,
\begin{eqnarray}
\alpha^{\rm M1}(^3\!P_2) \approx 2\,\alpha_0^{\rm M1}(^3\!P_2) \left(1 - \frac{m^2}{4} \right) .
\end{eqnarray}

\noindent If $m=2$, this contribution to $\alpha^{\rm M1}(^3\!P_2)$ becomes zero. If $m=0$, then we have
\begin{eqnarray}
\alpha^{\rm M1}(^3\!P_2) \approx 2\,\alpha_0^{\rm M1}(^3\!P_2) .
\label{al0M1}
\end{eqnarray}

Numerically, we find the matrix element $|\langle ^3\!P_1 ||\mu|| ^3\!P_2 \rangle|\approx 1.5811\,\mu_0$,
where $\mu_0 \approx 9.274 \times 10^{-24} \, {\rm J}/{\rm T}$ is the Bohr magneton, and note that the values obtained in the CI+MBPT and CI+all-order approximations begin to differ only in the 6{\it th} significant figure.\\

Using Eqs.~(\ref{DelE}) and (\ref{al0M1}) and the experimental differential energy $E_{^3\!P_2}-E_{^3\!P_1} \approx 124\,\, {\rm cm}^{-1}$, we arrive at
$$\Delta \nu^{(1)} \approx 18 \,\,\frac{\rm Hz}{({\rm mT})^2}\, B^2,$$ where the magnetic field is expressed in mT. This shift is slightly smaller than in the case of the $^{1}\mathrm{S}_0 \leftrightarrow \, ^{3}\mathrm{P}_{0}$ transition and can be readily accounted for by monitoring the magnetic field via the linear Zeeman shift of opposing hyperfine sublevels as has previously been done \cite{Bre18}.

\subsection{Electric Quadrupole Shift}
\subsubsection{Quadrupole moment}
The size of the electric quadrupole shift is determined by the quadrupole moment of the  $^3\!P_2$ state. In general, the quadrupole moment $\Theta$ of an atomic state $| J \rangle$ is given by
\begin{eqnarray}
\Theta &=& 2\, \langle J, M_J=J |Q_0| J, M_J=J \rangle \nonumber \\
       &=& 2\, \sqrt{\frac{J(2J-1)}{(2J+3)(J+1)(2J+1)}} \langle J ||Q|| J \rangle ,
\end{eqnarray}
where $ \langle J ||Q||  J \rangle$ is the reduced matrix element of the electric quadrupole operator. For the $3s3p\, ^3\!P_2$ state, we find
\begin{eqnarray}
\langle ^3\!P_2 ||Q|| ^3\!P_2 \rangle &\approx& 8.3\, e a_0^2
\label{Q_ME}
\end{eqnarray}
and
\begin{eqnarray}
\Theta (^3\!P_2) &\approx& 4.0\, e a_0^2 . 
\end{eqnarray}

For context, the quadrupole moment of the $^3\!P_2$ state in $^{27}$Al$^+$ is slightly larger than the quadrupole moments of the $^2D_{5/2}$ states in singly ionized alkaline earth elements \cite{quadrupole_moments}.

\section{Hyperfine Averaging}
As highlighted above, the angular momentum of the $J=2$, $^{3}\mathrm{P}_{2}$ state introduces larger Zeeman shifts and a larger electric quadrupole shift compared to the $^{1}\mathrm{S}_0$$\leftrightarrow$$^{3}\mathrm{P}_{0}$. Previously, Zeeman averaging of the $^{1}\mathrm{S}_0$$\leftrightarrow$$^{3}\mathrm{P}_{0}$ transition was performed by alternately driving $\Delta m_{F} = 0$ transitions from the two $m_{F} = \pm 5/2$ extreme states. The magnetic and electric quadrupole moments of the clock states in this transition come from the $^{27}$Al$^+$ nucleus so that the Zeeman shift and electric quadrupole shift are both small enough that the Zeeman averaging is needed only for monitoring the strength of the DC magnetic field. A slightly different averaging scheme is required for the $^{1}\mathrm{S}_0$$\leftrightarrow$$^{3}\mathrm{P}_{2}$ transition to eliminate the electric quadrupole shift and the leading-order Zeeman shifts, such as have been successfully applied in other clock species with $J>1/2$ states \cite{Barrett_field_insensitive, PTB_hyperfine_averagin, Huntemann10-18, Ca+LN2, Lu+2022comparison}.\\

It is well-known that the electric quadrupole shift and, in fact, all rank 2 tensor shifts can be canceled either by averaging over several $m_F$ levels for fixed $F$ or by averaging over several $F$ levels for fixed $m_F$ \cite{Itano_Hg+, Barrett_2015, Sr+_Quadrupole}. Both schemes are illustrated in Fig$.$\,\ref{fig:Al+ levels} with purple and green boxes, respectively. The latter scheme has been successfully demonstrated in $^{176}$Lu$^+$ and enables each component transition to be first-order magnetic field insensitive \cite{Barrett_2015}. Such a scheme would be well-suited to $^{26}$Al$^+$, which (although unstable) has even nuclear spin and several $m_F = 0$ levels. Here, we highlight the fixed $F$ averaging scheme which is conceptually very similar to the averaging scheme applied to the $^{1}\mathrm{S}_0$$\leftrightarrow$$^{3}\mathrm{P}_{0}$ clock transition detailed above. There, the average of the two transitions has no linear Zeeman shift, and the difference frequency is used to measure the magnetic field and constrain the quadratic Zeeman shift. Averaging over the two $^{3}\mathrm{P}_{2}$, $F=\frac{1}{2}$ states requires just as many transitions and eliminates the electric quadrupole shift. In both averaging schemes, rf or microwave drives, illustrated with black arrows in Fig$.$\,\ref{fig:Al+ levels}, applied during the spectroscopy sequence, could be used to eliminate the first-order sensitivity to magnetic field fluctuations \cite{Barrett_field_insensitive, PTB_hyperfine_averagin, Lu+2022comparison}. We note also that passive magnetic shielding is  used in other optical atomic clocks with higher sensitivity to first-order Zeeman shifts \cite{DubMadZho13, Ca+LN2}.

\begin{figure}
    \centering
    \includegraphics[width=\linewidth]{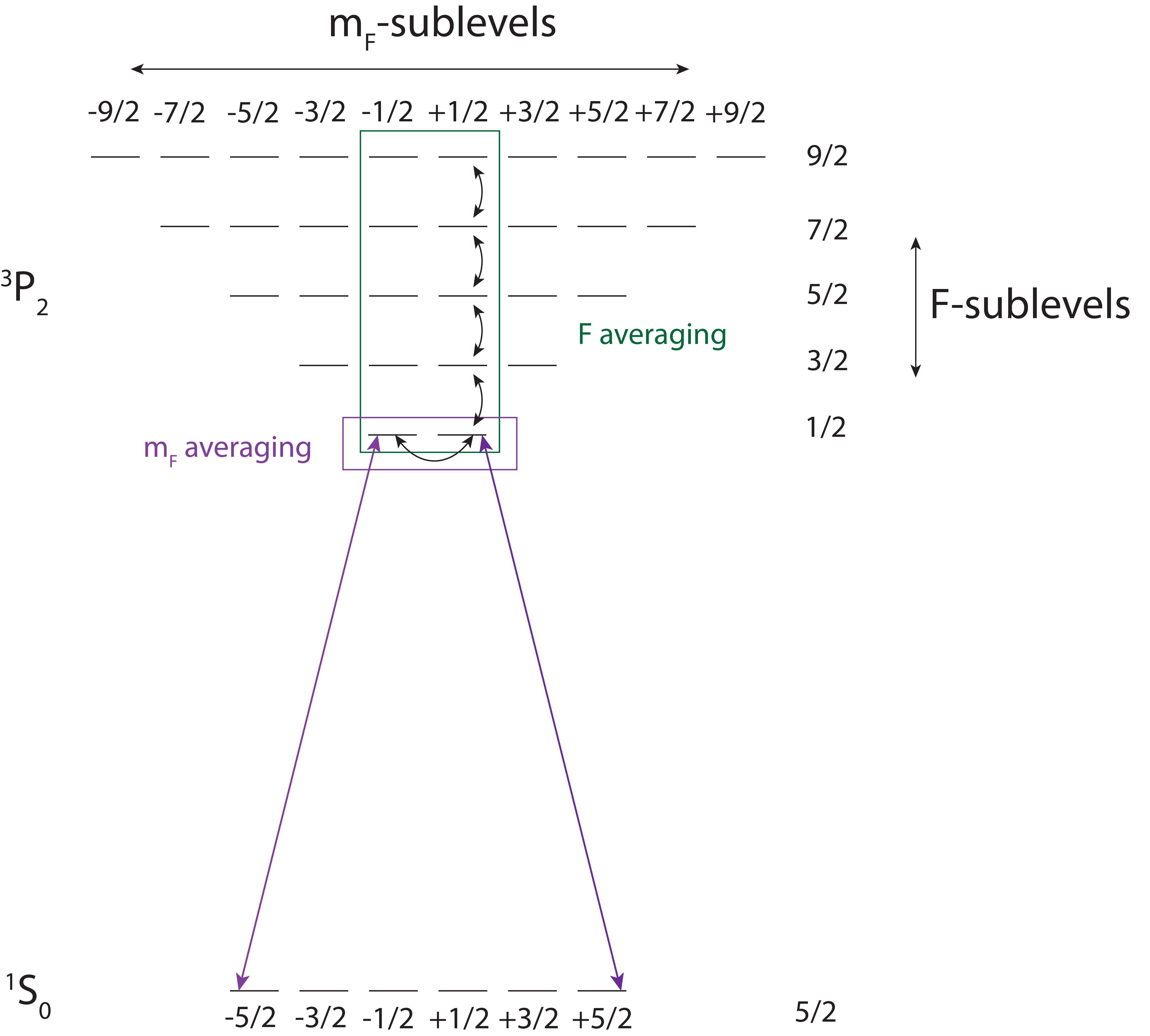}
    \caption{The hyperfine structure of the $^{1}\mathrm{S}_0$$\leftrightarrow$$^{3}\mathrm{P}_{2}$ transition in $^{27}$Al$^+$ is shown where the F-levels of the metastable $^{3}\mathrm{P}_{2}$ state are vertically offset and the $m_F$-levels are offset horizontally. Optical transitions from the ground state are illustrated in purple, while rf and microwave transitions are illustrated in black.}
    \label{fig:Al+ levels}
\end{figure}

\section{Proposed Measurements}
One of the strongest motivations for investigating the properties of the $^{1}\mathrm{S}_0$$\leftrightarrow$$^{3}\mathrm{P}_{2}$ transition in $^{27}$Al$^+$ is the longer lifetime of the excited state, around \SI{300}{\second} rather than \SI{20.7}{\second} \cite{RalKraRea11}. The standard quantum limit (SQL) imposes a strict limit on measurement stability during clock comparison measurements, and trapped-ion frequency standards using only single ions are often limited by probe durations. Techniques that allow clock comparison measurements beyond the laser coherence limit such as correlation spectroscopy recently demonstrated measurement stability with $^{27}$Al$^+$ consistent with the lifetime limit \cite{Cle20}, suggesting that further gains may be possible by using narrower transitions. Among other measurements, this increased stability could be used to measure differential gravitational redshifts at high precision or to measure the motional energy near the ground state via the time dilation shift, both of which would require months long measurement campaigns with the $^{1}\mathrm{S}_0$$\leftrightarrow$$^{3}\mathrm{P}_{0}$ transition.\\

However, it is not always the case that the $^{3}\mathrm{P}_{2}$ state has a longer lifetime than the $^{3}\mathrm{P}_{0}$ state in divalent species \cite{RalKraRea11}. As a result, although less likely to be directly applicable in $^{27}$Al$^+$ we highlight a few additional measurements that the auxilliary $^{1}\mathrm{S}_0$$\leftrightarrow$$^{3}\mathrm{P}_{2}$ transition makes possible that do not rely solely on the exact values of the lifetime.\\

Most generally, we note that an important use of a second, long-lived clock transition is to calibrate a primary clock transition. For example, although the electric quadrupole shift of the $^{1}\mathrm{S}_0$$\leftrightarrow$$^{3}\mathrm{P}_{0}$ transition in $^{27}$Al$^+$, due to the nuclear quadrupole moment, is around an order of magnitude below the current level of accuracy it could become a relevant systematic effect in future accuracy evaluations. The much larger electric quadrupole moment of the $^{3}\mathrm{P}_{2}$ state could then provide a means of measuring the electric quadrupole field, as is done with the E2 and E3 transitions in Yb$^+$. Similarly, interleaved measurements of individual hyperfine $^{1}\mathrm{S}_0$$\leftrightarrow$$^{3}\mathrm{P}_{2}$ transitions could potentially provide higher precision measurements of the DC magnetic field, and thus the residual quadratic Zeeman shift of the $^{1}\mathrm{S}_0$$\leftrightarrow$$^{3}\mathrm{P}_{0}$ transition. In this approach, the different sensitivities to various systematic effects of two clock transitions in a single species can be leveraged to constrain systematic shifts of the less sensitive transition. Of the examples highlighted here, magnetic field measurements with the $^{1}\mathrm{S}_0$$\leftrightarrow$$^{3}\mathrm{P}_{2}$ seems promising to improve the accuracy of the $^{1}\mathrm{S}_0$$\leftrightarrow$$^{3}\mathrm{P}_{0}$ transition.  

\subsection{Calibrating Clock Transitions: In Situ Measurement of the Blackbody Radiation Environment}
More novel measurements, however, are possible, and here we apply this approach to the case of measuring the BBR environment. Blackbody radiation is a prominent systematic shift in many optical atomic clocks that has remained difficult to accurately characterize. For example, characterization of the BBR shift in the Yb$^+$ ion clock \cite{Huntemann10-18} required detailed measurement and modelling of the BBR environment \cite{BBRYb+}, while a recent measurement campaign with two Lu$^+$ ion clocks refrained entirely from reporting the BBR shift of the clock transitions without verification from independent clock comparisons \cite{Lu+2022comparison}. In contrast, the most recent accuracy evaluations of the $^{27}\mathrm{Al}^+$ ion clock \cite{Bre18} and the $^{87}\mathrm{Sr}$ lattice clock \cite{Sr87-2015} were able to bound the BBR shift but unable to provide statistical errors. As a result, a direct method of measuring the equivalent temperature of the blackbody radiation environment, even if difficult, could be a valuable method to accurately measure and bound the BBR shift without modelling.\\

We reproduce calculations from Refs$.$\,\cite{RosenbandAl+BBR, Bre18} here, in which the Stark shift of an atomic state $\ket{a}$ in the presence of light at frequency $\omega$ and a electric field strength $E_0$ is given by

\begin{equation}
    \delta\nu_{a}\left(E, \omega\right) = -\frac{1}{4h}E^{2}\alpha_{a}\left(\omega\right),
\end{equation}
where the scalar polarizability $\alpha_{a}\left(\omega\right)$ is defined as,

\begin{equation}
    \alpha_{a}\left(\omega\right) = \frac{e^{2}}{m_{e}} \sum_{j} \frac{f_j}{\omega_{j}^{2} - \omega^{2}}.
\end{equation}

\noindent Here, $\omega_{i}$ and $f_{i}$ are the frequencies and oscillator strengths of all atomic transitions from $\ket{a}$, with the electron charge and mass denoted $e$ and $m_e$, respectively. The Stark shift, $\Delta\nu$, of the transition $\ket{a} \leftrightarrow \ket{b}$ is then

\begin{flalign}
    \Delta\nu_{a \rightarrow b}\left(E, \omega\right) &= \delta\nu_{b}\left(E, \omega\right) - \delta\nu_{a}\left(E, \omega\right)\\
    &= -\frac{1}{4h}E^{2}\Delta\alpha_{a \rightarrow b}\left(\omega\right),
\label{Eq:BBR Stark shift}
\end{flalign}

\noindent where $\Delta\alpha_{a \leftrightarrow b}\left(\omega\right)$ is the scalar differential polarizabiltiy listed in Table II for the $^{1}\mathrm{S}_0$$\leftrightarrow$$^{3}\mathrm{P}_{0}$ and $^{1}\mathrm{S}_0$$\leftrightarrow$$^{3}\mathrm{P}_{2}$ transitions, respectively. The Stark shift of a clock transition at frequency $\nu_{\mathrm{clock}}$ with differential polarizability $\Delta\alpha_{\mathrm{clock}}\left(\omega\right)$ in the presence of BBR at temperature $T$ is then,

\begin{equation}
    \Delta\nu_{\mathrm{clock}} = -\frac{1}{4\epsilon_{0}\pi^{3}c^{3}} \int_{0}^{\infty} \Delta\alpha_{\mathrm{clock}}\left(\omega\right) \frac{\omega^{3}}{e^{\hbar\omega/k_{B}T}-1} d\omega,
\end{equation}

which, when integrated, yields the well known scaling with $T^{4}$.\\

In general, the BBR environment ``seen'' by the clock ion is not well described by a single temperature. In the absence of a BBR shield of uniform temperature and emissivity \cite{YbLatticeBBRShield, SrLatticeRelativity, Ca+LN2}, the trap chamber is typically a polished metal well thermalized to the laboratory temperature, while some elements of the trap are locally heated, by up to a few degrees. We propose measuring the frequency difference of the $^{1}\mathrm{S}_0$$\leftrightarrow$$^{3}\mathrm{P}_{0}$ and $^{1}\mathrm{S}_0$$\leftrightarrow$$^{3}\mathrm{P}_{2}$ transitions, which we now denote as $\Delta\nu_{2,0}$, to high precision in a cryogenic environment with a blackbody radiation environment characterized by the the electric field contributing to the BBR shift, $\left< E^2 \right>_{4\mathrm{K}}$. The frequency difference, including the BBR shift of $\Delta\nu_{2,0}$, then becomes
\begin{equation}
    \Delta\nu_{4\mathrm{K}} = \left(\nu_2 - \nu_0\right) -\frac{1}{4h} \left< E^2 \right>_{4\mathrm{K}}\left( \Delta\alpha_{2} - \Delta\alpha_{0} \right)
\end{equation}
\noindent where $\nu_2$, $\nu_0$ and $\alpha_{2}$, $\alpha_{0}$ are the respective frequencies and differential polarizabilities of the $^{1}\mathrm{S}_0$$\leftrightarrow$$^{3}\mathrm{P}_{2}$ and $^{1}\mathrm{S}_0$$\leftrightarrow$$^{3}\mathrm{P}_{0}$ transitions. Near \SI{4}{\kelvin}, the BBR shift is suppressed by around eight orders of magnitude and we can safely make the approximation that $\Delta\nu_{2,0} \approx \Delta\nu_{4\mathrm{K}}$. When $\Delta\nu_{2,0}$ is again measured at room temperature, and even in an entirely different apparatus, the BBR shifted frequency difference is given by
\begin{equation}
    \Delta\nu_{300\mathrm{K}} = \left(\nu_2 - \nu_0\right) -\frac{1}{4h} \left< E^2 \right>_{300\mathrm{K}}\left( \Delta\alpha_{2} - \Delta\alpha_{0} \right).
\label{Eq: RT_BBR_shift}
\end{equation}
\noindent The differential polarizabilities $\Delta\alpha_{0}$, $\Delta\alpha_{2}$ can each be measured to high accuracy via a well characterized light shift on a co-trapped Ca$^{+}$ ion \cite{Barrett_polarizability}. The differential polarizability of the $^2\mathrm{S}_{1/2} \leftrightarrow$ $^2\mathrm{D}_{5/2}$ transition in $^{40}\mathrm{Ca}^{+}$ is around two orders of magnitude larger than both $\Delta\alpha_{0}$ and $\Delta\alpha_{2}$ and has been measured to a relative precision of around $3\times10^{-4}$ \cite{Ca+MagicRF}. The individual differential polarizabilities $\Delta\alpha_{0}$ and $\Delta\alpha_{2}$ can be measured in a similar manner by varying the intensity of laser coincident on the Ca$^{+}$ and Al$^{+}$ ions while monitoring the frequency shift of $\nu_{2}$ or $\nu_{0}$ against a stable reference. Eq$.$\,\ref{Eq: RT_BBR_shift} can be rearranged so that the electric field spectrum contributing to a BBR shift is given by,
\begin{equation}
    \left< E^2 \right>_{300\mathrm{K}} = 4h \frac{\Delta\nu_{4\mathrm{K}} - \Delta\nu_{300\mathrm{K}}}{\Delta\alpha_2 - \Delta\alpha_0}.
\end{equation}

Although this technique requires measurement in a cryogenic environment, we emphasize that this measurement only needs to be performed once and can be performed in a completely different apparatus from the one in which the clock, using either transition, will ultimately be operated in. The frequency difference $\nu_2 - \nu_0$ could be measured to high precision in a high-performance cryogenic apparatus located at a metrology institute which could then be used to calibrate the BBR environment of a higher-uncertainty room temperature system, e.g. a transportable clock. Transportable optical atomic clocks have been proposed for use in geodesy \cite{relativistic_geodesy, PTB_geodesy} and would naturally be exposed to various BBR environments, as a result, an in situ measurement could be crucial for low-uncertainty operation. Similarly, as a transportable optical atomic clock is likely to use a lower stability probe laser, the measurement would benefit from spectroscopy schemes such, as correlation spectroscopy, which enable high-stability frequency difference measurements beyond the limit imposed by laser phase noise \cite{Cle20}. Such a technique could be easily applied to two co-trapped ions as demonstrated in Ref$.$\,\cite{Chou_3ion}.\\

Nevertheless, the insensitivity to BBR of both the $^{1}\mathrm{S}_0$$\leftrightarrow$$^{3}\mathrm{P}_{0}$ and $^{1}\mathrm{S}_0$$\leftrightarrow$$^{3}\mathrm{P}_{2}$ transitions means that such a measurement would inevitably be difficult and require averaging down to below the \SI{1e-18}{} level to provide errors that are similar to the non-statistical errors reported in Ref$.$\,\cite{Bre18}. However, we note that the $^{3}\mathrm{P}_{0}$ lifetime limited stability of correlation spectroscopy \cite{Cle20} allows a fractional statistical uncertainty of \SI{1e-18}{} to be reached with less than one day of measurement time.

\subsection{Searching for New Physics: Violation of Local Lorentz Invariance}
Finally, we investigate the suitability of the $^{1}\mathrm{S}_0$$\leftrightarrow$$^{3}\mathrm{P}_{2}$ transition in the search for physics beyond the Standard Model. Two clock transitions in a single species can be a valuable tool here, with the most well known example being the Yb$^+$ ion, where a comparison of an electric quadrupole and electric octupole transition currently set the tightest bounds on potential time variation of the fine structure constant \cite{Lange_alphadot}. Similarly, an apparent oscillation of these transition energies would be a signature of ultralight dark matter \cite{clock_DM, BACON}. In the case of $^{27}\mathrm{Al}^{+}$, as a low-$Z$, light ion, sensitivity to variation of the fine structure constant is minimal for both the $^{1}\mathrm{S}_0$$\leftrightarrow$$^{3}\mathrm{P}_{0}$ and $^{1}\mathrm{S}_0$$\leftrightarrow$$^{3}\mathrm{P}_{2}$ transitions. However, as the $^{3}\mathrm{P}_{2}$ state has non-zero orbital angular momentum it is sensitive to potential violation of local Lorentz invariance (LLI) and Einstein's equivalence principle (EEP). Here we calculate these sensitivities and note a similar consideration applies to many other optical atomic clocks based on divalent atoms and ions.\\

Violation of local Lorentz invariance (LLI) and Einstein's equivalence principle (EEP) in bound electronic states results in a small shift of the Hamiltonian that can be described by~\cite{HohLeeBud13}
\begin{equation}
\delta H=-\left(  C_{0}^{(0)}-\frac{2U}{3c^{2}}c_{00}\right)
\frac{\mathbf{p}^{2}}{2}-\frac{1}{6}C_{0}^{(2)}T^{(2)}_{0},\label{eq1}%
\end{equation}
where we use atomic units, $\mathbf{p}$ is the momentum of a bound electron, $c$ is the speed of
light, and
\begin{equation*}
T^{(2)}_{0}\equiv\mathbf{p}^{2}-3p_{z}^{2}.
\end{equation*}
The coupling constants $C_{0}^{(q)},U,$ and $c_{00}$ are discussed in detail in~\cite{HohLeeBud13}.\\

The change in $^{27}\mathrm{Al}^{+}$ energy levels depends on the values of
the $\langle J m |\mathbf{p}^{2}| J m \rangle$ and $\langle J m |\mathbf{p}^{2}-3p_{z}^{2}| J m \rangle$ matrix elements. The stretched and reduced matrix elements of the $T^{(2)}_0$ operator are connected as
\begin{equation*}
\langle J m |T^{(2)}_0|J m \rangle=(-1)^{j-m}\left(
\begin{array}
[c]{ccc}%
 J   & 2 &  J \\
-m & 0 & m
\end{array}
\right)  \langle J ||T^{(2)}|| J \rangle.
\end{equation*}

\noindent Writing down the $3j$-symbol in explicit form, we have,
\begin{eqnarray}
\langle J m |T^{(2)}_0 |J m \rangle &=& \frac{-J(J+1) + 3m^2}{\sqrt{(2J+3) (J+1) (2J+1) J (2J-1)}} \nonumber \\
&\times& \langle J ||T^{(2)}|| J \rangle.
\label{eq10}
\end{eqnarray}
The value of the angular factor in Eq.~(\ref{eq10}) for $^3\!P_2$ is $-0.2390 + 0.1195\,m^2$.

Calculations were carried out in the CI, CI+MBPT, and CI+all-order approximations. The RPA corrections,
which describe a reaction of the core electrons to an externally applied perturbation, were included.
The results are listed in Table~\ref{Tab:LLI} in atomic units. We note that we present the \textit{reduced} matrix elements
for the $T^{(2)}_0$ operator but the stretched matrix elements for $\mathbf{p}^{2}$ because this is a scalar operator.
\begin{table}
\caption{\label{Tab:LLI} The results obtained in the CI, CI+MBPT, and CI+all-order approximations,
including the RPA corrections, (in a.u.)}
\begin{ruledtabular}
\begin{tabular}{lcccc}
            Matrix element                               &  CI   & CI + MBPT & CI + All \\
\hline \\ [-0.6pc]
$\langle ^1\!S_0  |\mathbf{p}^2| ^1\!S_0 \rangle$        & 3.39  &  3.55     &  3.55    \\[0.2pc]
$\langle ^3\!P_2  |\mathbf{p}^2| ^3\!P_2 \rangle$        & 3.05  &  3.17     &  3.17    \\[0.2pc]
$\langle ^3\!P_2  |\mathbf{p}^2| ^3\!P_2 \rangle$ -
$\langle ^1\!S_0  |\mathbf{p}^2| ^1\!S_0 \rangle$        &-0.34  & -0.37     & -0.38    \\[0.4pc]
\hline \\ [-0.6pc]
$\langle ^3\!P_2||T^{(2)}||^3\!P_2 \rangle$              & 3.58  &  3.70     &  3.70
\end{tabular}
\end{ruledtabular}
\end{table}

We use the CI + all-order values given in the last column of Table~\ref{Tab:LLI} for the matrix elements to obtain the energy shift.
Atomic units are converted to Hz using $1$ a.u. $\approx\left(  6.57968\times10^{15}\text{ Hz}\right)  h$.

We obtain the following.%
\begin{eqnarray*}
\langle ^1\!S_0 |\mathbf{p}^2 |^1\!S_0 \rangle & \approx & 3.55\, {\rm a.u.} \approx (2.3 \times 10^{16}\, {\rm Hz})\, h , \nonumber \\
\langle ^3\!P_2 |\mathbf{p}^2 |^3\!P_2 \rangle & \approx & 3.17\, {\rm a.u.} \approx (2.1 \times 10^{16}\, {\rm Hz})\, h , \nonumber \\
\langle ^3\!P_2 |T^{(2)}_0| ^3\!P_2 \rangle & =& ( -0.884+0.442\,m^2 )\times 3.70 \text{ a.u.} \nonumber \\
 &\approx& \left[ (-2.2  + 1.1\, m^2) \times 10^{16}\, \text{Hz} \right] h , \\
\end{eqnarray*}
Substituting these values into Eq.~(\ref{eq1}), we obtain the frequency shift (in Hz)%
\begin{widetext}
\begin{eqnarray*}
^1\!S_0:  && \frac{\Delta E}{h} \approx -1.2 \times 10^{16} \left( C_0^{(0)} - \frac{2U}{3c^2} c_{00} \right) , \\
^3\!P_2:  && \frac{\Delta E}{h} \approx -1.1 \times 10^{16} \left( C_0^{(0)} - \frac{2U}{3c^2} c_{00} \right)
                                        + ( 3.7 - 1.8 \, m^{2}) \times10^{15}\, C_0^{(2)}.
\end{eqnarray*}
\end{widetext}

The LLI-induced energy shift between the highest ($m=2$) and lowest ($m=0$) sublevels of the $^3\!P_2$ state is given by
\begin{equation}
 \left| \frac{\Delta E}{h\,C_0^{(2)}} \right| \approx 7.2 \times 10^{15}\, {\rm Hz} .
\end{equation}

For comparison, this sensitivity is around 50\% larger than in Ca$^+$, which previously was used to set the tightest constrains on LLI \cite{Haeffner_LLI_with_entanglement}, and around an order of magnitude smaller than Yb$^{+}$, which provides the current best constraints \cite{Yb+_LLI}. Again, given that $^{27}\mathrm{Al}^{+}$ is a low-$Z$, relatively non-relativistic ion, this result is somewhat surprising and suggests that future investigation into other species is fruitful. An immediate example is the $^{115}\mathrm{In}^{+}$ ion clock with $Z=49$, with a similarly unused $^{1}\mathrm{S}_0$$\leftrightarrow$$^{3}\mathrm{P}_{2}$ transition lying far in the UV but still laser accessible. Optical clocks based on Pb$^{2+}$ \cite{pb2+} ($Z = 82$) and Sn$^{2+}$ \cite{sn2+_arxiv} ($Z = 50$) may be promising systems for LLI searches in the future. Notably, lead and tin have many stable, even isotopes and the $^{1}\mathrm{S}_0$$\leftrightarrow$$^{3}\mathrm{P}_{2}$ transition, in combination with the $^{1}\mathrm{S}_0$$\leftrightarrow$$^{3}\mathrm{P}_{0}$, would be a natural candidate for King plot nonlinearity measurements \cite{king_plot_1, king_plot_2}.\\

\section{Outlook}
Optical atomic clocks based on a $^{1}\mathrm{S}_0$$\leftrightarrow$$^{3}\mathrm{P}_{0}$ transition are both common and highly successful including, among others, clocks using neutral Sr or Yb and trapped ion clocks using Al$^{+}$ and In$^{+}$. Universally, these species also contain an electric quadrupole $^{1}\mathrm{S}_0$$\leftrightarrow$$^{3}\mathrm{P}_{2}$ transition that is typically unused. We have investigated the properties of this transition in $^{27}\mathrm{Al}^{+}$ and found that it retains many favorable properties of the $^{1}\mathrm{S}_0$$\leftrightarrow$$^{3}\mathrm{P}_{0}$ transition, most notably the low sensitivity to BBR. While the nonzero orbital angular momentum of this state makes the spectroscopy sequence slightly more complex, we show that many of the techniques used in other clock species with similar $J > 0$ states are easily applicable and that the use of transitions involving $J > 0$ states does not impact fundamental accuracy or stability.\\

We have also described a few example use cases of this transition, for example an in-situ method of measuring both the BBR coefficient and equivalent electric field intensity at the position of a trapped ion - a measurement that has been difficult to characterize and bound but which can provide valuable input to evaluate systematic uncertainties and to test atomic structure calculations. We also highlight the potential of the under-utilized $^{1}\mathrm{S}_0$$\leftrightarrow$$^{3}\mathrm{P}_{2}$ transition to provide constraints on physics beyond the Standard Model - in the case of $^{27}\mathrm{Al}^{+}$, a search for LLI and EEP violation. Here, we note that calculation of the clock-related atomic properties of the $^{1}\mathrm{S}_0$$\leftrightarrow$$^{3}\mathrm{P}_{2}$ transition in other established clock systems seems particularly fruitful. In species with many stable, even, isotopes e.g. Yb and Sn$^{2+}$, the existence of a second clock transition could prove beneficial for King-plot nonlinearity tests and other searches for new physics. Meanwhile, in systems such as $^{27}\mathrm{Al}^{+}$ where the lifetime of the $^{3}\mathrm{P}_{0}$ state is limited by hyperfine mixing, the longer-lived $^{3}\mathrm{P}_{2}$ state could potentially be used in high-stability measurements in which the atomic state lifetime is a limiting factor \cite{Cle20}. Notably, the atomic state lifetime is already a limitation for the In$^{+}$ ion clock while the roughly \SI{300}{\second} lifetime of the $^{3}\mathrm{P}_{2}$ state in $^{27}\mathrm{Al}^{+}$ balances the possibility of long probe durations with low probe light induced Stark shifts. 

\section{Acknowledgements}
We thank W.F. McGrew and Y.S. Hassan for for careful reading and their suggestions on the manuscript.
 This work was supported by the National Institute of Standards and Technology, NSF Quantum Leap Challenge Institute Award OMA - 2016244, the Office of Naval Research (Grant Number N00014-20-1-2513), and the European Research Council (ERC) under the European Union's Horizon 2020 research and innovation program (Grant Number 856415). This research was supported in part through the use of University of
Delaware HPC Caviness and DARWIN computing systems: DARWIN - A Resource for Computational and Data-intensive Research at the University of Delaware
and in the Delaware Region, Rudolf Eigenmann, Benjamin E. Bagozzi, Arthi Jayaraman, William Totten, and Cathy H. Wu, University of Delaware, 2021, URL:
https://udspace.udel.edu/handle/19716/29071.

\clearpage
\pagebreak

\section{Appendix}
\subsection{Hyperfine Formalism}
We use the basis set $|Jm I \lambda \rangle = |Jm\rangle |I \lambda \rangle$, where ${\bf F} = {\bf J} + {\bf I}$ and $M = m + \lambda$.
Keeping only the magnetic-field dipole interaction in the operator $H_{\rm hfs}$ , we have
\begin{eqnarray}
 H &=& H_{\rm hfs} - {\boldsymbol \mu}_{\rm at} {\bf B} \nonumber \\
   &=& h A_{\mathrm{hfs}}\, {\bf J}{\bf I} + (\mu_0 g_J {\bf J} + \mu_N g_I {\bf I})\,{\bf B}.
\label{A:H}
\end{eqnarray}

Assuming that $\bf B$ is directed along the $z$ axis and neglecting the third term in~\eref{A:H}
(because $\mu_0 \gg \mu_N$) we can write \eref{A:H} as
\begin{eqnarray}
 H &\approx& h A_{\mathrm{hfs}} I_z J_z + \frac{hA_{\mathrm{hfs}}}{2} (J_+ I_- + J_- I_+) + \mu_0 g_J J_z B \nonumber \\
   &=& (h A_{\mathrm{hfs}} I_z + \mu_0 g_J B) J_z + \frac{hA_{\mathrm{hfs}}}{2} (J_+ I_- + J_- I_+).
\label{A:Hapr}
\end{eqnarray}

Here, $J_{\pm}$ and $I_{\pm}$ are the ladder operators for which
\begin{eqnarray}
 J_{\pm}|J m \rangle &=& \sqrt{(J \mp m)(J \pm m+1)} |J, m+1\rangle , \nonumber \\
 I_{\pm}|J \lambda \rangle &=& \sqrt{(I \mp \lambda)(I \pm \lambda+1)} |I, \lambda+1\rangle .
\label{A:lad}
\end{eqnarray}

Using Eqs.~(\ref{A:Hapr}) and (\ref{A:lad}) we can express
$\langle J m' I \lambda' |H| J m I \lambda \rangle$ as
\begin{eqnarray}
&&\langle J m' I \lambda' | (hA_{\mathrm{hfs}} I_z + \mu_0 B g_J)\,J_z | J m I \lambda \rangle \nonumber \\
&=& \delta_{m m'} \delta_{\lambda \lambda'} (hA_{\mathrm{hfs}} \lambda + \mu_0 B g_J) m .
\end{eqnarray}

Using \eref{A:lad} we find
\begin{eqnarray*}
 \langle J m' |J_+| J m \rangle &=& \sqrt{(J - m)(J + m+1)} \langle m',m+1 \rangle , \nonumber \\
  &=& \delta_{m',m+1} \sqrt{(J - m)(J+m+1)} ,
\label{}
\end{eqnarray*}
and
\begin{eqnarray*}
 \langle I \lambda' |I_+| I \lambda \rangle &=& \sqrt{(I - \lambda)(I + \lambda+1)} \langle \lambda',\lambda+1 \rangle , \nonumber \\
  &=& \delta_{\lambda',\lambda+1} \sqrt{(I - \lambda)(I+\lambda+1)} .
\label{}
\end{eqnarray*}

Then
\begin{eqnarray}
&&\frac{hA_{\mathrm{hfs}}}{2}\langle J m' I \lambda' | J_+ I_- | J m I \lambda \rangle = \delta_{m',m+1} \delta_{\lambda',\lambda-1} \nonumber \\
&\times&  \sqrt{(J - m)(J + m+1)(I + \lambda)(I - \lambda+1)}
\end{eqnarray}
and
\begin{eqnarray}
&&\frac{hA_{\mathrm{hfs}}}{2}\langle J m' I \lambda' | J_- I_+ | J m I \lambda \rangle = \delta_{m',m-1} \delta_{\lambda',\lambda+1} \nonumber \\
&\times&  \sqrt{(J + m)(J - m+1)(I - \lambda)(I + \lambda+1)}.
\end{eqnarray}

In total, we obtain
\begin{widetext}
\begin{eqnarray}
\langle J m' I \lambda' |H| J m I \lambda \rangle
 &=& \delta_{m m'} \delta_{\lambda \lambda'} (hA_{\mathrm{hfs}} \lambda + \mu_0 B g_J) m \nonumber \\
 &+& \frac{hA_{\mathrm{hfs}}}{2} \left[ \delta_{m',m+1} \delta_{\lambda',\lambda-1}\sqrt{(J - m)(J + m+1)(I + \lambda)(I - \lambda+1)} \right. \nonumber \\
 && \hspace{1.2pc} +\left. \delta_{m',m-1} \delta_{\lambda',\lambda+1}\sqrt{(J + m)(J - m+1)(I - \lambda)(I + \lambda+1)}\right].
\label{A:lam}
\end{eqnarray}

Taking into account that $\lambda = M-m$, we can rewrite the equation above as follows.
\begin{eqnarray}
\langle J m' I (M-m') |H| J m I (M-m) \rangle
 &=& \delta_{m m'} \left\{hA_{\mathrm{hfs}} (M-m) + \mu_0 B g_J\right\}\,m \nonumber \\
 &+& \frac{hA_{\mathrm{hfs}}}{2} \left[ \delta_{m',m+1} \sqrt{(J - m)(J + m+1)(I + M-m)(I - M+m+1)} \right. \nonumber \\
 && \hspace{1.2pc} +\left. \delta_{m',m-1} \sqrt{(J + m)(J - m+1)(I - M+m)(I + M-m+1)}\right].
\label{A:HM}
\end{eqnarray}
\end{widetext}
\subsection{Application of the formalism to the $^3\!P_2$ state.}
Here, we apply this formalism to the $^3\!P_2$ state. Because $J$ and $I$ are the same for an initial and final state,
we use a shorter notation for the basis states $|m,(M-m)\rangle$ instead of $|J m, I (M-m)\rangle$.

We have $J=2$ and, respectively, $m=-2,-1,0,1,2$. Then we can construct the $5 \times 5$ matrix
$\langle m',(M-m') |H| m,(M-m) \rangle$. As follows from \eref{A:HM}, only elements of the main and two
secondary diagonals of this matrix will be non-zero. Designating the matrix elements by $a_{ij}$
we obtain the following non-zero elements.
\begin{widetext}
\begin{eqnarray}
a_{11} &\equiv& \langle 2,(M-2) |H| 2,(M-2) \rangle = 2[hA_{\mathrm{hfs}}(M-2) + \mu_0 g_J B], \nonumber \\
a_{12} = a_{21} &\equiv& \langle 2,(M-2) |H| 1,(M-1) \rangle = hA_{\mathrm{hfs}}\sqrt{(I+M-1)(I-M+2)}, \nonumber \\
a_{22} &\equiv& \langle 1,(M-1) |H| 1,(M-1) \rangle = hA_{\mathrm{hfs}}(M-1) + \mu_0 g_J B, \nonumber \\
a_{23} = a_{32} &\equiv& \langle 1,(M-1) |H| 0,M \rangle = hA_{\mathrm{hfs}}\sqrt{\frac{3}{2}(I+M)(I-M+1)}, \nonumber \\
a_{34} = a_{43} &\equiv& \langle 0,M |H| -1,M+1 \rangle = hA_{\mathrm{hfs}}\sqrt{\frac{3}{2}(I+M+1)(I-M)}, \nonumber \\
a_{44} &\equiv& \langle -1,(M+1) |H| -1,(M+1) \rangle = -hA_{\mathrm{hfs}}(M-2) - \mu_0 g_J B, \nonumber \\
a_{45} = a_{54} &\equiv& \langle -1,(M+1) |H| -2,(M+2) \rangle = hA_{\mathrm{hfs}}\sqrt{(I+M+2)(I-M-1)}, \nonumber \\
a_{55} &\equiv& \langle -2,(M+2) |H| -2,(M+2) \rangle = -2[hA_{\mathrm{hfs}}(M+2) + \mu_0 g_J B].
\label{A:Ha}
\end{eqnarray}
\end{widetext}
Let us consider two particular cases of $M=9/2$ and $M=7/2$.
\subsubsection{$\bf M=9/2$}
We fix $M=9/2$. There is only one value of the total angular momentum, $F=9/2$, which corresponds to $M=9/2$.
Since $M=m+\lambda$, for $M=9/2$ there is only one option: $m=2$ and $\lambda=5/2$.
Then, from \eref{A:lam} we obtain
\begin{eqnarray}
 \langle 2,5/2 |H| 2,5/2 \rangle = 5hA_{\mathrm{hfs}} + 2\mu_0 g_J B.
\end{eqnarray}
As we can see, in this case, there is no term $\sim B^2$.
\subsubsection{$\bf M=7/2$}
Now, fix $M=7/2$. There are two values $F=9/2$ and $F=7/2$, the projection of which can be equal to $M=7/2$.
It can be obtained in only two ways: $|m=2, \lambda=3/2 \rangle$ and $|m=1, \lambda=5/2 \rangle$
We can define these states as a basis. To find the energy shift, $\Delta E$, we have to solve the equation
\begin{equation}
 \langle \Psi | H |\Psi \rangle = \Delta E ,
\end{equation}
where
\begin{equation}
 |\Psi \rangle =
     \begin{pmatrix}
       |m=2, \lambda=3/2 \rangle \\
       |m=1, \lambda=5/2 \rangle
     \end{pmatrix} .
\end{equation}

Using \eref{A:Ha} and noting that only $a_{ij}$ ($i,j=1,2$) will be not equal to zero, we obtain
\begin{widetext}
\begin{eqnarray}
a_{11} &\equiv& \langle 2,3/2 |H| 2,3/2 \rangle = 3hA_{\mathrm{hfs}} + 2\mu_0 g_J B, \nonumber \\
a_{12} = a_{21} &\equiv& \langle 2,3/2 |H| 1,5/2 \rangle = hA_{\mathrm{hfs}}\sqrt{(I+5/2)(I-3/2)} = \sqrt{5}\,hA_{\mathrm{hfs}}, \nonumber \\
a_{22} &\equiv& \langle 1,5/2 |H| 1,5/2 \rangle = \frac{5}{2}hA_{\mathrm{hfs}} + \mu_0 g_J B ,
\end{eqnarray}
where we accounted for $I=5/2$.
\end{widetext}
Thus, we have to solve the equation
\begin{equation*}
\left(
\begin{array}{cc}
3hA_{\mathrm{hfs}} + 2\mu_0 g_J B - \Delta E &        \sqrt{5}\,hA_{\mathrm{hfs}} \\
       \sqrt{5}\,hA_{\mathrm{hfs}}           & \frac{5}{2}hA_{\mathrm{hfs}} + \mu_0 g_J B - \Delta E
 \end{array}
\right) = 0.
\end{equation*}

The solutions to this equation are as follows.
\begin{eqnarray}
&&\Delta E_{1,2} = \frac{11 h A_{\mathrm{hfs}}}{4} + \frac{3}{2}\, \mu_0 g_J B \nonumber \\
&\pm& \frac{9 h A_{\mathrm{hfs}}}{4} \sqrt{ 1 + \frac{4 \mu_0 g_J B}{81\,h A_{\mathrm{hfs}}} + \frac{4\, (\mu_0 g_J B)^2}{81 (h A_{\mathrm{hfs}})^2} } .
\label{A:DelE12}
\end{eqnarray}

Then, extracting the quadratic contribution from $B$, designated as $\Delta E^{(2)}$, we find
\begin{eqnarray*}
\left|\Delta E^{(2)}\right| = \frac{1}{18}\,\frac{(\mu_0 g_J)^2}{h A_{\mathrm{hfs}}}\,B^2 .
\label{x2}
\end{eqnarray*}

To verify \eref{A:DelE12} we can put $B=0$, arriving at
\begin{eqnarray}
 \Delta E_1 = 5 h A_{\mathrm{hfs}}, \nonumber \\
 \Delta E_2 = \frac{1}{2} h A_{\mathrm{hfs}} .
\end{eqnarray}

From this we obtain
\begin{equation}
 \Delta E_1 - \Delta E_2 = (9/2)hA_{\mathrm{hfs}},
\end{equation}
 as it should be, because $\Delta E_1 = \Delta E_{F=9/2}$, $\Delta E_2 = \Delta E_{F=7/2}$ and
\begin{equation}
 \Delta E_F - \Delta E_{F-1} = FhA_{\mathrm{hfs}} .
\end{equation}

\bibliography{AlII}
\end{document}